\documentclass[superscriptaddress]{article}
\usepackage{graphicx}
\usepackage{amssymb}
\usepackage{hyperref}
\usepackage[utf8]{inputenc}
\usepackage{csquotes}
\usepackage{amsmath}
\usepackage{bm}
\usepackage{color}
\usepackage[english]{babel}
\usepackage[backend=biber,sorting=none]{biblatex}
\usepackage{authblk}
\title{Topological Floquet interface states\\ in optical fibre loops}
\author[1]{A. Bisianov}
\author[1]{A. Muniz}
\author[1]{U. Peschel}
\author[1]{O. A. Egorov}
\affil[1]{ Institute of Solid State Theory and Optics, Friedrich Schiller University Jena, Max-Wien-Platz 1, 07743 Jena, Germany}
\date{\today}                     
\begin{document}
\maketitle

\begin{abstract}
We experimentally observe a coexisting pair of topological anomalous Floquet interface states in a (1+1)-dimensional Discrete Photon Walk. We explicitly verify the robustness of these states against local static perturbations respecting chiral symmetry of the system, as well as their vulnerability against non-stationary perturbations. The walk is implemented based on pulses propagating in a pair of coupled fibre loops of dissimilar lengths with dynamically variable mutual coupling. The topological interface is created via phase modulation in one of the loops, which allows for an anomalous Floquet topological transition at the interface. 
\end{abstract}
\maketitle

\section{Introduction}
It is known from condensed matter physics \cite{Kane05b,Kane05a,Bernevig06} that topologically protected interface and edge states can guide light and particle waves in a robust and scattering-free way. Moreover, symmetry-protected topological states are considered as a potentially very powerful basis for fault-tolerant quantum computations \cite{Else16,Prakash15,Raussendorf17}. Meanwhile, these states have been discovered in many optical structures embracing photonic lattices \cite{Ganainy15,Rechtsman13,Wang09,Hromada09}, metamaterials \cite{Khanikaev13}, coupled resonator arrays \cite{Hafezi13,Hafezi11}, quasi-crystals \cite{Verbin13,Verbin15,Kraus12} and Photonic Quantum Walks \cite{Kitagawa12,Chen18,Wang18}. 

The presence of non-trivial topological order/phases of gapped Hamiltonians is predetermined by their underlying fundamental symmetries \cite{Kitagawa10a,Schnyder08,Roy17}. In this study, we are particularly interested in one of such symmetries, the so-called chiral symmetry, also known as sublattice symmetry. The simplest one-dimensional system, that obeys chiral symmetry and at the same time supports non-trivial topological order is the well-known Su-Schrieffer-Heeger (SSH) model \cite{Su79}, originally proposed to model the dynamics of non-interacting electrons in a one-dimensional two-partite chain. Here, sublattice symmetry ensures topological protection of localized states, which appear at interfaces (edges) between a topologically non-trivial domain of the SSH two-partite chain and a trivial one (vacuum).  The topological protection means that isolated localized states must be spectrally located in the middle of the energy band gap (zero-energy modes) and therefore never couple to extended bulk modes as long as the gap remains open and no strong perturbations violate chiral symmetry. The fundamental principle standing behind the topological protection is known as the bulk-edge correspondence \cite{Asboth13}. It states that midgap localized states, existing at an interface/edge of the bulk, are predefined by topological invariants of the neighboring bulk domains, which are global characteristics of the bulk’s topology, and do not depend on the exact form of the edge/interface or any local perturbations imposed in this region. The important clause, however, is that neither the edge/interface itself nor potential external perturbations are allowed to break the underlying symmetry of the system. Provided that chiral symmetry is intrinsically local, i.e. it can be defined individually for each lattice site or dimer, the latter condition can be typically fulfilled for a wide variety of edges/interfaces and local disorders, thus ensuring the very existence of topological states as well as their topological protection against the latter ones.

The topological order becomes much more interesting and richer, when it comes to periodically driven topological insulators, which are typically treated in the framework of effective Floquet (stroboscopic) Hamiltonians. The first observation was that such dynamical systems may feature edge or interface states even if the topological invariants are zero and therefore indicate no topological phase transition. Such anomalous states have been experimentally observed in many two-dimensional systems \cite{Wang18,Mukherjee17,Maczewsky17,Leykam16,Peng16} as well as in one-dimensional chiral ones, including a periodically driven SSH model \cite{Cheng19} and a Discrete Quantum Walk \cite{Kitagawa12}. Later, it was realized that the conventional topological invariants, which are winding numbers in 1D \cite{Kitagawa10a,Asboth14,Asboth13,Obuse15} and Chern numbers in 2D \cite{Rudner13,Nathan15,Kitagawa10b}, have to be appropriately modified and supplemented, in order to unambiguously predict the number of topological states and thus to establish the bulk-edge correspondence. It is worth mentioning, that in the special case of chiral one-dimensional Floquet insulators as well as in many other classes the bulk-edge correspondence had been strictly proven only for bulks with translational symmetry, as it was still based on winding numbers. However, recent theoretical advances \cite{Liu18} rigorously show that a generalization of the winding number to inhomogeneous systems is possible, implying that the bulk-edge correspondence remains in force also for inhomogeneous systems and systems with disorder. 

The ever-growing excitement about anomalous Floquet states has initiated many new frontiers of topology, such as the concept of “non-Hermitian Boundary State Engineering” \cite{Lee16,Hoeckendorf19,Li19}, enhanced quantum computing via Floquet topological states \cite{Bomantara18,Zhou18} and non-adiabatic quantized charge pumping \cite{Titum16}. In this paper, we will experimentally demonstrate a coexisting pair of anomalous Floquet interface states, which naturally arise in the context of recent studies \cite{Bisianov19,Pankov19} on the fully temporal analogue of the Discrete Photonic Quantum Walk \cite{Kitagawa12}. Furthermore, we will experimentally reveal a decisive role of chiral symmetry for their topological robustness since in 1D this aspect, supporting the above mentioned bulk-edge correspondence for disordered Floquet systems \cite{Liu18}, has been to date lacking an experimental approval. In particular, we will show that initially topologically protected interface states may leave their midgapped energy positions and may even finally dissolve in the bulk if the symmetry is locally broken.

\begin{figure}
\includegraphics[width=\textwidth]{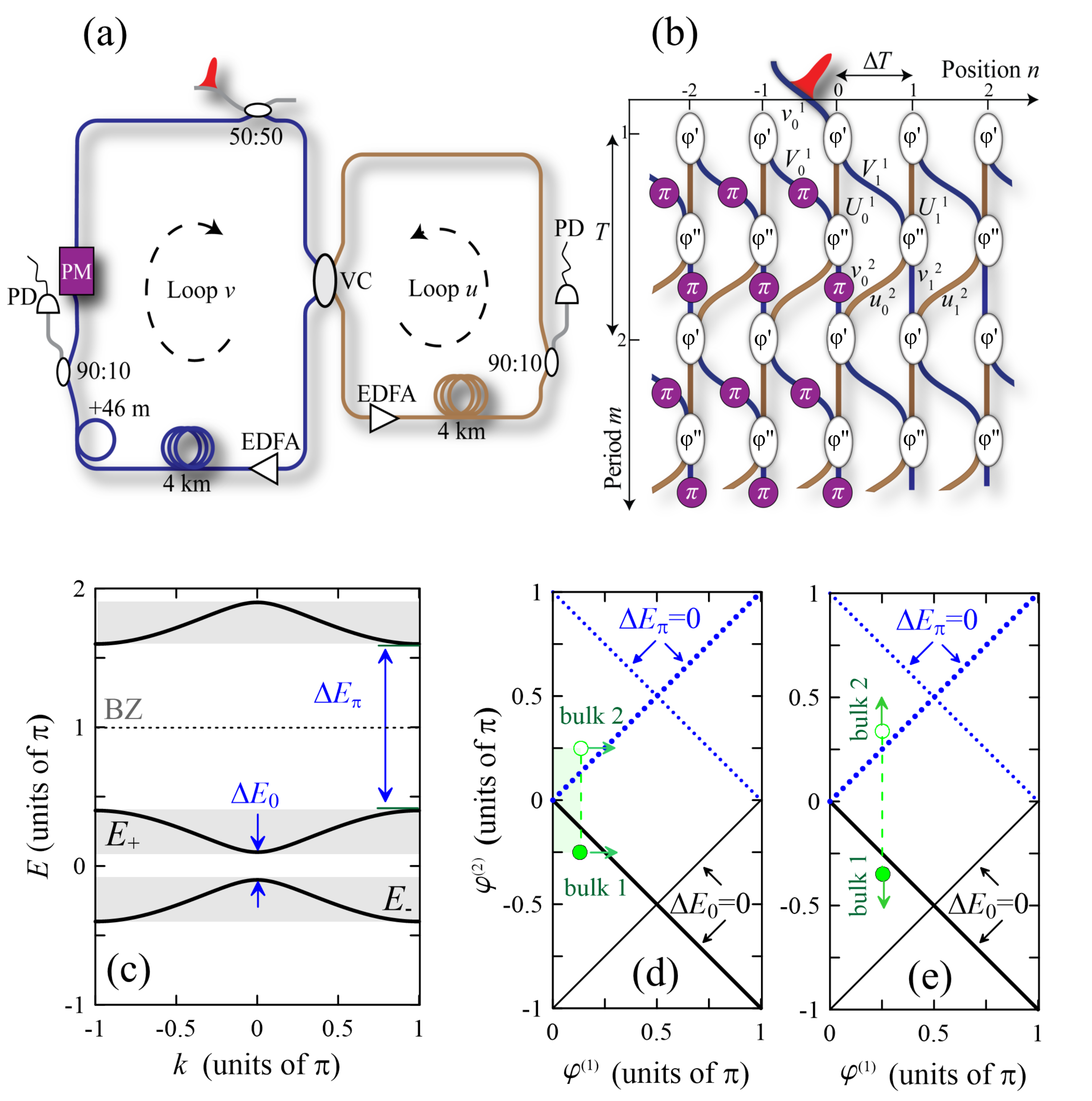}
\caption{(a) Our experimental set-up consists of two fibre loops with the mutual length difference $\Delta L$, which are coupled by a voltage-controlled variable coupler (VC). Pulses are launched into and extracted from the loops by two optical beam splitters. Pulse train evolution is monitored with photo diodes (PD), one per loop. Erbium-doped fibre amplifiers (EDFA) compensate for all losses. (b) Pulse evolution in the fibre loops projected onto a synthetic spatiotemporal lattice. (c) Band structure of bulk waves [see Eq.\:(9)] for ${{\varphi }^{(1)}}=\pi /4$ and ${{\varphi }^{(2)}}=-0.35\pi $. A dashed horizontal line marks the border of the Brillouin zone (BZ). Band gaps form at $\Delta {{E}_{0}}$ and $\Delta {{E}_{\pi }}$. (d) and (e) Phase diagrams showing gap closings on the parameter plane  . The black thick (thin) lines depict the gap closure $\Delta {{E}_{0}}$ for $k = 0$ ($k  = \pi$). The blue dotted thick (thin) lines mark the gap closure $\Delta {{E}_{\pi }}$ for $k = \pi$ ($k = 0$). The vertical dashed line in panel (d) indicates the first configuration with topological transition between bulk 1 with ${{\varphi }^{(1)}}=\varphi $, ${{\varphi }^{(2)}}=-\pi /4$ and bulk 2 with ${{\varphi }^{(1)}}=\varphi $, ${{\varphi }^{(2)}}=\pi /4$, where $\varphi >0$ is a free parameter. The second configuration in panel (e) indicates the transition between bulk 1 with ${{\varphi }^{(1)}}=\pi /4$, ${{\varphi }^{(2)}}=-\varphi $ and bulk 2 with ${{\varphi }^{(1)}}=\pi /4$, ${{\varphi }^{(2)}}=\varphi $.}
\label{Fig1_Sketch}
\end{figure}

\section{The experimental set-up and governing \\equations}
Our experimental set-up consists of a pair of coupled fibre loops, whose lengths have an average value of about $L\approx4100$ metres and differ by $\Delta L\approx46$ metres. 25 nanosecond pulses with carrier wavelength of 1552 nanometres are injected into the longer loop by a fibre coupler, the open end of which is used to constantly monitor the field evolution inside the loops by photo receivers [see Fig.\:\ref{Fig1_Sketch}(a)]. Once launched a pulse travels through the fibre loops and splits every round trip into two replicas entering different loops. Owing to the non-zero length difference between the loops, propagating pulses acquire path dependent delays. During the round trips they, therefore, populate a discrete lattice of equally spaced time slots. Our system has a generic two round trip periodicity and we follow this periodicity while modulating the fibre loop set-up. Hence, one can distinguish two principal time scales, namely the duration of two successive round-trips of the pulses through the short and long loops $T = 2L/c \approx 41.4$ µs ($c \approx 2\times10^8$ m/s is the speed of pulses in the fibre) and the temporal shift, which pulses acquire after one round trip when passing to the longer instead of the shorter loop $\Delta T = \Delta L/c \approx 230$ ns. In what follows, we formally link the number of double round trips with a discrete number of time steps $m$, while the smaller delay $\Delta T$ becomes associated with a discrete transverse position $n$. Thus, we project the monitored pulses onto a synthetic lattice with one temporal and one pseudo-spatial dimension [see Fig.\:\ref{Fig1_Sketch}(b)]. Dispersive spreading of individual nanosecond pulses is insignificant for the given propagation lengths, implying that every pulse circulating in the short and the long loop can be fully described by the complex amplitudes $u_n^m$ and $v_n^m$, respectively. The coupling ratio $t/r$ between the loops ($t^2$ defines the portion of the pulse energy, which stays in the same loop and $r^2$ that switching to the other loop, so that $t^2 + r^2 = 1$) can be dynamically tuned by an optical voltage-controlled variable coupler with a characteristic response time (rise and fall) of 250 ns and a bandwidth of 100 kHz. In addition, we modulate the phase in the long loop $v$ using a dynamically controlled electro-optic phase modulator. 

The iterative evolution of the pulse amplitudes during two successive round trips is explicitly described by

\begin{eqnarray}
U_{n}^{m}&=&\cos \phi _{n}^{\prime}\ u_{n}^{m}+i\sin \phi _{n}^{\prime}\,\ v_{n}^{m}e^{  i\alpha _{n}^{\prime} },\\                                       \label{eq_model_U}
V_{n}^{m}&=&\cos \phi _{n-1}^{\prime}\ \,v_{n-1}^{m}e^{ i\alpha _{n-1}^{\prime} }+i\sin \phi _{n-1}^{\prime}\,\ u_{n-1}^{m},\\ \label{eq_model_V}
u_{n}^{m+1}&=&\cos \phi _{n+1}^{\prime\prime}\ U_{n+1}^{m}+i\sin \phi _{n+1}^{\prime\prime}\,\ V_{n+1}^{m}e^{  i\alpha _{n+1}^{\prime\prime} },\\ \label{eq_model_u}
v_{n}^{m+1}&=&\cos \phi _{n}^{\prime\prime}\ \,V_{n}^{m}e^{  i\alpha _{n}^{\prime\prime} }+i\sin \phi _{n}^{\prime\prime}\ \,U_{n}^{m} \label{eq_model_v}
\end{eqnarray}
where $U_{n}^{m}$ and $V_{n}^{m}$ denote the amplitudes at the intermediate time step recorded after one round trip through the short loop ${{T}_{\text{short}}}=(L-0.5\Delta L)/c$. 

In a stationary case, for which the system does not change between successive time steps $m$, that is coupling parameters $\phi_{n}^{\prime}$, $\phi _{n}^{\prime\prime}$ and variations $\alpha _{n}^{\prime}$, $\alpha _{n}^{\prime\prime}$ have no $m$-dependence, a so-called Floquet ansatz can be applied to generate solutions of the form $\left| \Psi _{n}^{m} \right\rangle \equiv \left| {{u}_{n}},{{v}_{n}},{{U}_{n}},{{V}_{n}} \right\rangle \exp \left( -iEm \right)$ with a quasi-energy $E$ in the interval $-\pi \le E\le \pi $. 

In what follows, we restrict ourselves to the phase modulations that are multiples of $\pi$. In that case the system shows a kind of chiral symmetry with a spectrum being mirror symmetric around zero energy. Hence, for each eigenstate $|{{u}_{n}},{{v}_{n}},{{U}_{n}},{{V}_{n}}\rangle $ with an energy $E$ exists a mirror state $|u_{n}^{*},-v_{n}^{*},U_{n}^{*},-V_{n}^{*}\rangle $ with the opposite energy $-E$. As nontrivial localized states are non-degenerate eigenstate with quasi-energy 0 or $\pi$ each of them can be mapped onto itself based on the above symmetry relation. Hence its individual components ${{u}_{n}}$ and ${{v}_{n}}$ must be real except multiplication with a trivial phase (for our notation $i$ in case of ${{v}_{n}}$). 

We have preliminarily checked that the relative phase between the complex transmission elements of the coupler is always constant with a value of $\pi/2$. Hence, without additional measures we have only access to real valued and positive transmission and reflection coefficients with values between 0 and 1. This imposes a strong parametric limitation of $0\le \phi _{n}^{\prime}, \phi _{n}^{\prime\prime}\le \pi /2$, which does not allow one to observe anomalous topological states [see Figs.\:\ref{Fig1_Sketch}(d) and (e)]. But, by applying the phase shift of $\pi$ in the long loop, we circumvent this constraint. In what follows, we assume stationary conditions (no $m$-dependence) and apply phase shifts of zero or $\pi$ only ($\alpha _{n}^{\prime}=\alpha _{n}^{\prime\prime}=0$ or $\pi $). In that case, the original system (1 - 4) is equivalent to the following effective model:

\begin{eqnarray}
{U}_{n}^{m}&=&\cos \varphi _{n}^{(1)}\tilde{u}_{n}^{m}+i\sin \varphi _{n}^{(1)}\,\tilde{v}_{n}^{m},\\                                       \label{eq_model_UU}
{V}_{n}^{m}&=&\cos \varphi _{n-1}^{(1)}\,\tilde{v}_{n-1}^{m}+i\sin \varphi _{n-1}^{(1)}\,\tilde{u}_{n-1}^{m},\\ \label{eq_model_VV}
\tilde{u}_{n}^{m+1}&=&\cos \varphi _{n+1}^{(2)}{U}_{n+1}^{m}+i\sin \varphi _{n+1}^{(2)}\,{V}_{n+1}^{m},\\ \label{eq_model_uu}
\tilde{v}_{n}^{ m+1}&=&\cos \varphi _{n}^{(2)}\,{V}_{n}^{m}+i\sin \varphi _{n}^{(2)}\,{U}_{n}^{m} \label{eq_model_vv}
\end{eqnarray}
for which the new variables are given by $\varphi _{n}^{(1)}={{\phi }_{n}^{\prime}}$, $\varphi _{n}^{(2)}=\exp \left( i{{\alpha }_{n}} \right){{\phi }_{n}^{\prime\prime}}$, $\tilde{u}_{n}^{m}=u_{n}^{m}$ and $\tilde{v}_{n}^{m}=\exp \left( i{{\alpha }_{n}} \right)v_{n}^{m}$, where by definition $\alpha _{n}=\alpha _{n}^{\prime}=\alpha _{n}^{\prime\prime}$. This effective change of the sign of the reflection coefficient $r_{n}^{m}$ allows us to access the full range of topological phases, to be discussed further on.

\section{Floquet-Bloch waves}
Before addressing states at interfaces we study bulk states (no $n$-dependence of system parameters). We apply the Floquet-Bloch ansatz and seek for spatially extended plane wave solutions, having the form of $\left| \Psi _{n}^{m} \right\rangle \equiv \left| u\left( k \right),v\left( k \right) \right\rangle e^{\left(kn-E\left( k \right)m \right)i}$ and evolving stroboscopically in time. Owing to the spatial periodicity of the walk, also wave numbers $k$ of the Floquet-Bloch modes are limited to the first Brillouin zone, namely $-\pi<k<\pi$. 

Substituting this ansatz into equations (5-8) and solving a standard eigenvalue problem with respect to $E$ yields the following band structure
\begin{eqnarray}
{{E}_{\pm }}=\pm \text{arccos}(\cos {{\varphi }^{(1)}}\cos {{\varphi }^{(2)}}\cos k-\sin {{\varphi }^{(1)}}\sin {{\varphi }^{(2)}}),\label{eq_BandStruct}
\end{eqnarray}
which is symmetric with respect to zero energy as expected from the symmetry considerations made above.

The two dispersion branches $E_{\pm}(k)$ predefine the bulk properties of the Bloch-Floquet waves inside the first Brillouin zone (BZ), where gaps $\Delta E_0$ and $\Delta E_\pi$ are found around quasi-energies $E = 0$ and $E = \pm\pi$, respectively [see Fig.\:\ref{Fig1_Sketch}(c)]. Depending on the coupling angles ${{\varphi }^{(1)}}$ and ${{\varphi }^{(2)}}$, gaps may close thus leading to a sudden change of the topological phase. For arguments unambiguously linking the band gap closings with topological transitions, the reader may refer to a series of previous studies devoted to Photonic Quantum Walks \cite{Kitagawa12,Kitagawa10a,Bisianov19,Asboth12,Obuse15}. Straightforward arithmetic calculations based on Eq.\:(\ref{eq_BandStruct}) define lines of gap closures displayed in Figs.\:\ref{Fig1_Sketch}(d) and (e). In particular, the gap at $E = 0$ disappears given that ${{\varphi }^{(1)}}+{{\varphi }^{(2)}}=\pi p$ or ${{\varphi }^{(1)}}-{{\varphi }^{(2)}}=\pi /2+\pi p$ is fulfilled, with $p$ being an integer. In contrast, the gap at $E = \pm\pi$ closes for ${{\varphi }^{(1)}}-{{\varphi }^{(2)}}=\pi p$ or ${{\varphi }^{(1)}}+{{\varphi }^{(2)}}=\pi /2+\pi p$.

\section{Topologically protected interface states}
Let us consider an interface at $n = 0$ between a left-hand (bulk 1) and right-hand (bulk 2) sided domains, introduced via the phase shift $\alpha _{n}^{{}}=\pi $ for $n\le 0$ and 0, otherwise, which is equivalent to flipping the sign of the parameter ${{\varphi }^{(2)}}$ within the effective model (5-8). More precisely, we are going to study topologically protected interface states forming at the interface between two areas with opposite signs of coupling in every second round trip. 

Based on the bulk-boundary correspondence, the existence of interface states is expected as a result of a topological transition between regions with different topological phases characterized by the respective topological invariant, the Zak phase, or by its counterpart, the Winding number. Contrary to the SSH model, where zero (non-zero) Zak phase always implies a topologically trivial (non-trivial) state, in the periodically driven system under consideration such a relationship is not unambiguous anymore. Even for a zero Zak phase, the Floquet lattice may nevertheless support topologically non-trivial interface states, which typically emerge in pairs \cite{Kitagawa12,Kitagawa10a,Kitagawa10b}.

Here we use another approach \cite{Asboth12}, which also yields the number of existing interface states based on the parameters characterizing the bulk states on both sides of the interface \cite{Kitagawa12,Kitagawa10a,Bisianov19}. As one example, we select for bulk 1 (left side of the interface) ${{\varphi }^{(1)}}=\varphi $ and ${{\varphi }^{(2)}}=-\pi /4$ with $\varphi >0$ being a free parameter, while bulk 2 (right side) is defined by the parameters ${{\varphi }^{(1)}}=\varphi $ and ${{\varphi }^{(2)}}=\pi /4$. When connecting these two parameter points with a straight line as illustrated in Fig.\:\ref{Fig1_Sketch}(d) we count the parity of gap closures in the first Brillouine zone, thus obtaining the invariants $Q_0$ and $Q_\pi$ for the gap $\Delta {{E}_{0}}$ and $\Delta {{E}_{\pi }}$, respectively. Both invariants $Q_0$ and $Q_\pi$ are equal to 1 in the interval $0<\varphi <\pi /4$ corresponding to two interface states having quasi-energies $E = 0$ and $E = \pi$, respectively. Again, we would like to emphasize that contrariwise standard calculations \cite{Bisianov19} reveal zero Zak phase for both bulk domains, thereby indicating no phase transition and hence no localized states at the interface.

\begin{figure}
\includegraphics[width=\textwidth]{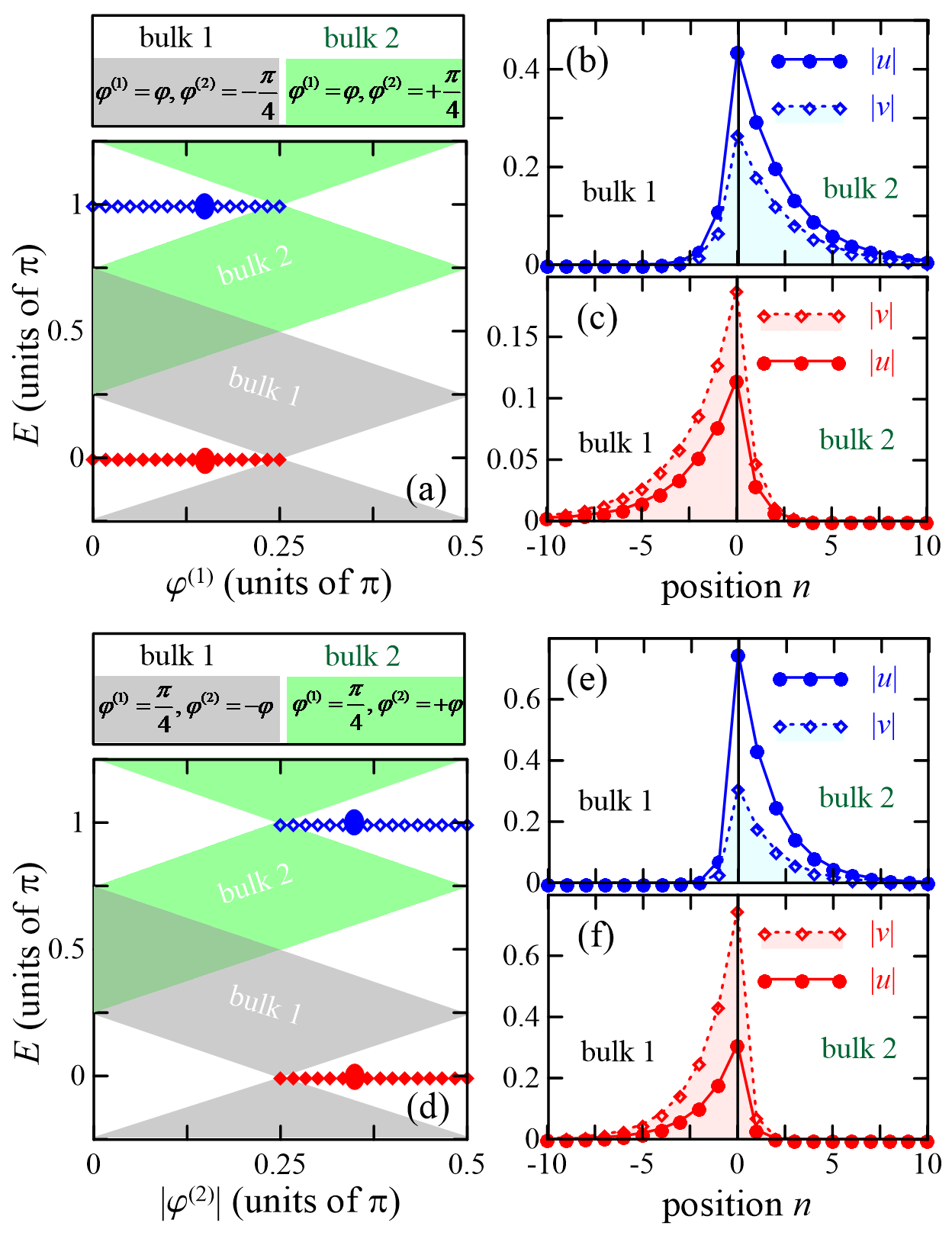}
\caption{(a) Calculated quasi-energies of the linear extended (gray and green areas) and the topologically protected (blue and red dots) interface states (bulk 1: ${{\varphi }^{(1)}}=\varphi $ and ${{\varphi }^{(2)}}=-\pi /4$, bulk 2: ${{\varphi }^{(1)}}=\varphi $ and ${{\varphi }^{(2)}}=\pi /4$, as sketched in Fig.\:\ref{Fig1_Sketch}(d)). (b), (c) Spatial profiles of the interface modes in gap $\Delta {{E}_{\text{ }\!\!\pi\!\!\text{ }}}$  (b) and $\Delta {{E}_{0}}$  (c) calculated for ${{\varphi }^{(1)}}=0.15\pi $. (d) Calculated quasi-energies of the bulk (gray and green areas) and the topologically protected localized (blue and red dots) interface states (bulk 1: ${{\varphi }^{(1)}}=\pi /4$ and ${{\varphi }^{(2)}}=\varphi $, bulk 2: ${{\varphi }^{(1)}}=\pi /4$ and ${{\varphi }^{(2)}}=-\varphi $, as shown in Fig.\:1 (e). Spatial profiles of the interface modes in gap $\Delta {{E}_{\text{ }\!\!\pi\!\!\text{ }}}$ (e) and $\Delta {{E}_{0}}$ (f) calculated for $\left| {{\varphi }^{(2)}} \right|=0.35\pi $.}
\label{Fig2_InterfaceModes}
\end{figure}

In what follows, we consider two principal cases, namely the previously discussed one $0<{{\varphi }^{(1)}}<{\pi }/{4}, \left| {{\varphi }^{(2)}} \right|={\pi }/{4}$, and ${{\varphi }^{(1)}}={\pi }/{4}\;,\ \left|  {{\varphi }^{(2)}} \right|>{\pi }/{4}\;$, presented in Fig.\:\ref{Fig1_Sketch}(e). In Figs.\:\ref{Fig2_InterfaceModes}(a) and (d), we explicitly verify the existence of the localized states by numerically calculating the band structures for the two principal cases. Typical amplitude profiles of the steady-state solutions are shown in Figs.\:\ref{Fig2_InterfaceModes}(b,c,e,f), featuring exponentially decaying tails with a main spreading into opposite directions of the interface for the 0- and $\pi-$energy states. It comes as no surprise, since for the chosen parameters the 0 gap in bulk 1 is almost closed while the same is the case for the $\pi$ gap in bulk 2. Consequently, the zero-energy state [Figs.\:\ref{Fig2_InterfaceModes}(c,f)] is almost delocalized in bulk 1, contrary to the $\pi-$energy state, [Figs.\:\ref{Fig2_InterfaceModes}(b,e)], which extends deep into bulk 2. 
In order to double-check the existence of the topologically protected interface states we investigated a particular choice of parameters, namely ${{\varphi }^{(1)}}={\pi }/{4}\;,\ \left|  {{\varphi }^{(2)}} \right|={\pi }/{2}\;$, for which the system decays into an ensemble of separated double strands and can be described analytically. As expected, two interface states with eigenvalues $E=0$ and $E=\pi$ are found, the field of which are non-zero at the boundary only, where they have the form ${{v}_{0}}=i\left( \sqrt{2}+1 \right){{u}_{0}}$ and ${{v}_{0}}=i\left( -\sqrt{2}+1 \right){{u}_{0}}$, respectively. Due to the mirror symmetry of the spectrum these singular eigenvalues cannot move, even for parameter variations. As already predicted above these eigenstates are symmetry protected and must exist in the whole parameter space, in which respective gaps remain open and which also covers the experimentally accessed parameter range [see Fig.\:\ref{Fig1_Sketch}(e)]. 

\begin{figure}
\includegraphics[width=\textwidth]{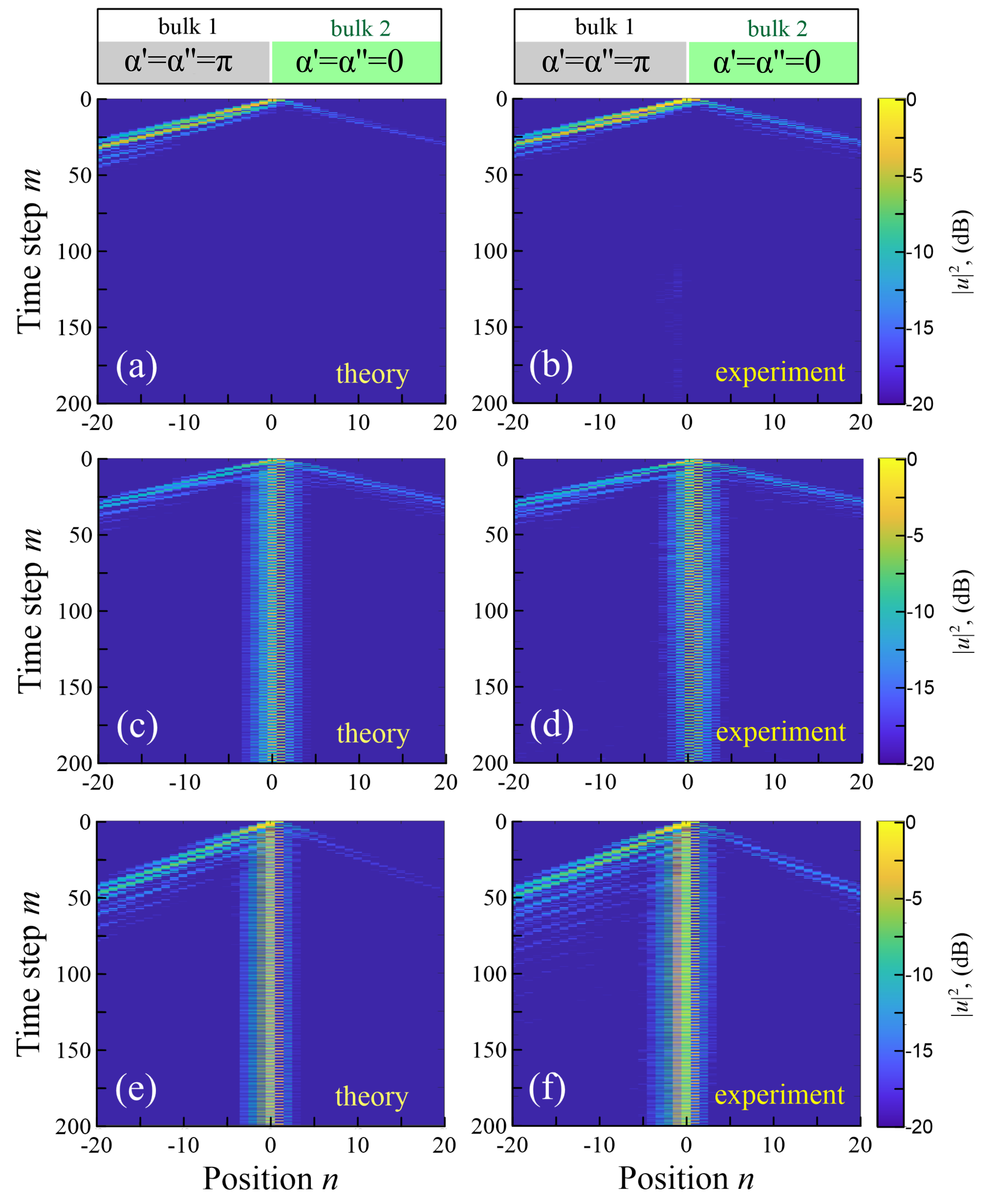}
\caption{(a), (c), (e) Numerical and (b), (d), (f) experimental dynamics of on-site excitation of the long loop $v$ in different regimes. The pulse propagation is traced here only in the short loop $u$. (a), (b) The interface states are not allowed for $\phi^{\prime} =0.3\pi $ and $\phi^{\prime\prime} =\pi /4$. (c),(d) Excitation of the couple of interface states for $\phi^{\prime} =0.1\pi $ and $\phi^{\prime\prime} =\pi /4$  [the case shown in Fig.\:\ref{Fig1_Sketch}(d) and Fig.\:\ref{Fig2_InterfaceModes}(a)]. (e), (f) Excitation of the couple of interface states for $\phi^{\prime} =\pi /4$ and $\phi^{\prime\prime} =0.35\pi $  [the case in Fig.\:\ref{Fig1_Sketch}(e) and Fig.\:\ref{Fig2_InterfaceModes}(d)]. The numerical modelling was performed within the original model (1-4) with the phase modulation of the long loop $\alpha _{n}=\pi $ within the bulk 1 (on the left-hand side).}
\label{Fig3_IntfcModesPropagation}
\end{figure}

\section{Experiments on the excitation and beating dynamics of interface states}
Interface states are excited by injecting a single pulse into the longer loop $v$ at $n=0$ as demonstrated by numerical and experimental measurements displayed in Figs.\:\ref{Fig3_IntfcModesPropagation}. In the topologically trivial case with no gap closing between the different touching bulk systems, the initial pulse energy transforms completely into propagating volume waves [see Figs.\:\ref{Fig3_IntfcModesPropagation}(a,b)]. In contrast, we demonstrate in Figs.\:\ref{Fig3_IntfcModesPropagation}(c,d) and Figs.\:\ref{Fig3_IntfcModesPropagation}(e,f) the existence of non-trivial interface states for the configurations considered in Fig.\:\ref{Fig2_InterfaceModes}(a) and \ref{Fig2_InterfaceModes}(d), respectively. The experimentally obtained excitation efficiency is critically dependent on the degree of spatial localization of the particular interface state, since the input field distribution is a single pulse in our experiments, which does not have a perfect overlap with an interface state. As we inject directly at the interface position into the longer loop $v$, typically the zero-energy state, which is mainly located in the left bulk [see Fig.\:\ref{Fig2_InterfaceModes}] is excited stronger than the $\pi$ state resulting in an asymmetric energy distribution [see Figs.\:\ref{Fig3_IntfcModesPropagation}(c,d,e,f)]. In addition, some energy is radiated away from the interface as propagating waves.
\begin{figure}
\includegraphics[width=\textwidth]{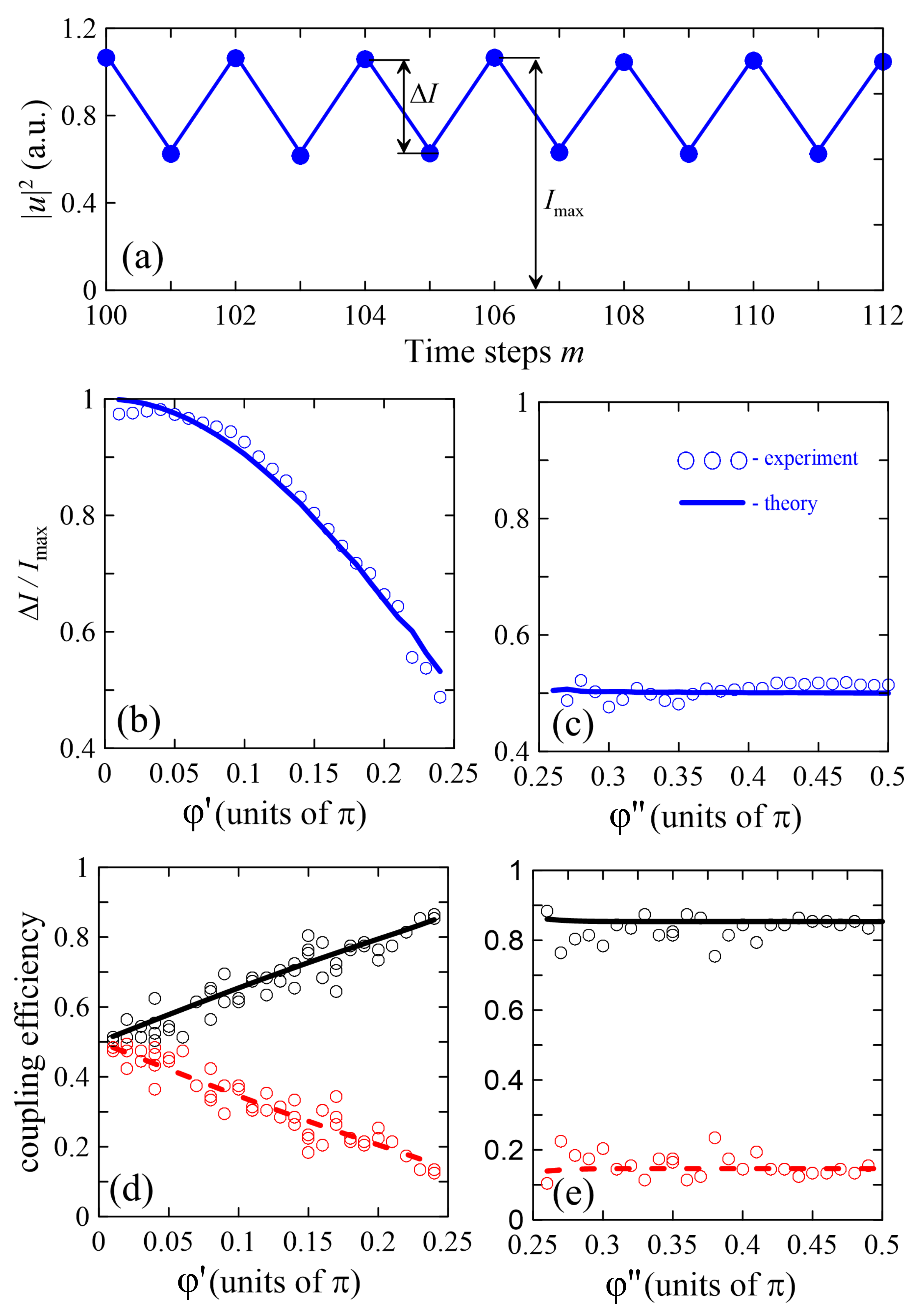}
\caption{(a) Temporal evolution of the intensity in the short loop at the interface position ($n=0$), measured experimentally at $\phi^{\prime} =\pi /4$ and $\phi^{\prime\prime} =0.35\pi $ (b) Relative beating amplitudes versus $\phi^{\prime} $ for the fixed value $\phi^{\prime\prime} =\pi /4$. (c) Relative beating amplitudes versus $\phi^{\prime\prime} $ for the fixed value $\phi^{\prime} =\pi /4$. (d),(e) Power efficiency of the coupling to the interface modes for the cases presented in (b) and (c), respectively. The black solid lines show numerically defined energy weight of the state at $E=0$, whereas the dashed red lines correspond to the interface state at $E=\pi $. Open circles show the energy weights derived from experimental measurements via the beating-based technique, explained in Appendix A. }
\label{Fig4_ModeBeating}
\end{figure}

Being simultaneously excited, the two topological Floquet states perform a fast power beating with a period of four round trips ($\Delta m=2$), since their quasi-energies differ by $\Delta E=\pi $ [see Fig.\:\ref{Fig4_ModeBeating}(a)]. Respective experimentally determined beating amplitudes are high and close to theoretical values [see Figs.\:\ref{Fig4_ModeBeating}(b) and (c)]. Based on these measurements we can reconstruct the relative energy distributions and deduce the coupling efficiencies [see Appendix A for details on the reconstruction technique], which agree well with theoretical predictions [see Figs.\:\ref{Fig4_ModeBeating}(d)]. Those depend critically on the varied parameter $\phi^{\prime}$  [see Figs.\:\ref{Fig4_ModeBeating}(d) and compare with Fig.\:\ref{Fig2_InterfaceModes}(a)]. Since the component $|v|$ of the surface mode within the gap $E=0$ is larger, its excitation by an external input beam launched into the long loop $v$ becomes more effective. In the second configuration for a fixed $\phi^{\prime}=\pi /4$ (equivalent to ${{\varphi }^{(1)}}=\pi /4$), the coupling efficiencies remain almost unaffected by the variation of $\phi^{\prime\prime} $, which is associated with $\left| {{\varphi }^{(2)}} \right|$ [see Fig.\:\ref{Fig4_ModeBeating}(e) and compare with Fig.\:\ref{Fig2_InterfaceModes}(d)]. 

\begin{figure}
\includegraphics[width=\textwidth]{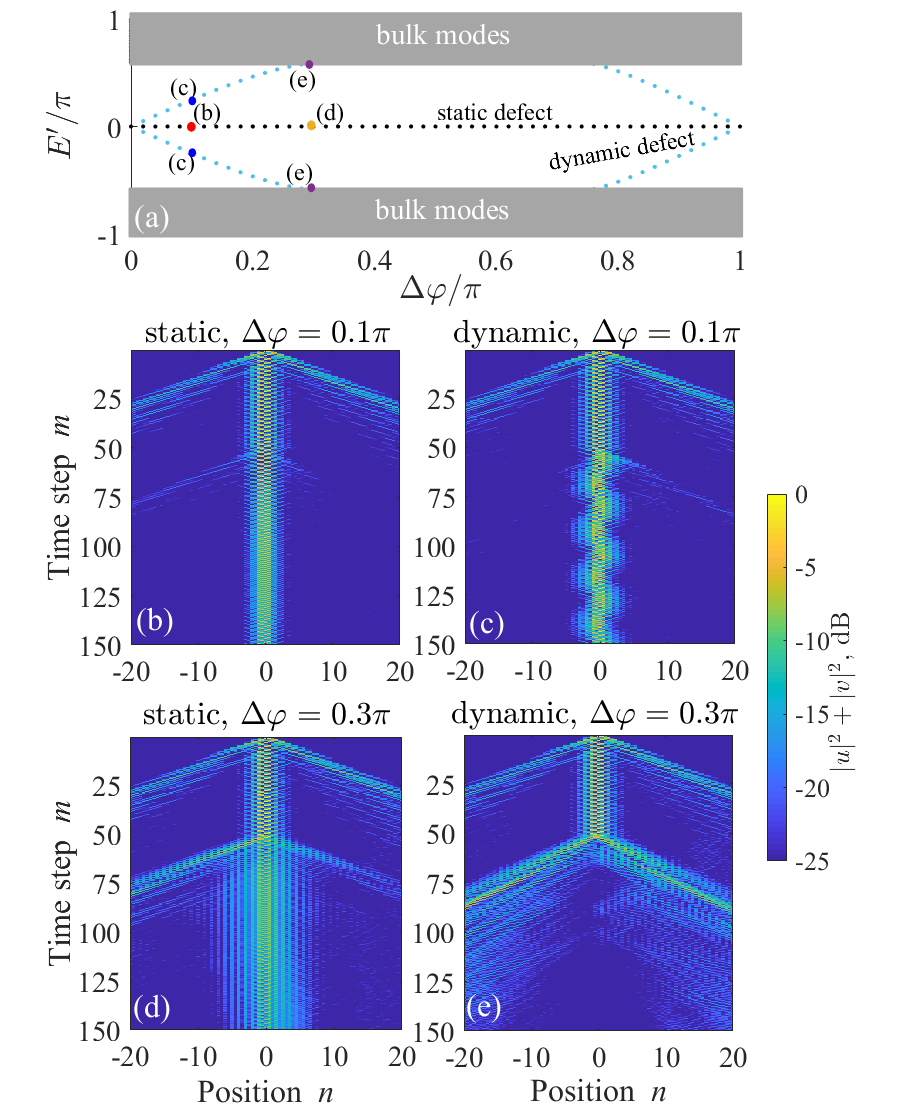}
\caption{(a) The back-folded spectrum of the bulk modes and the interface states in the presence of the static $\phi_{0}^{m}=\phi^{\prime} +\Delta \varphi $ (black dots) and the doubly periodic dynamic $\phi _{0}^{m}=\phi^{\prime}+{{\left( -1 \right)}^{m}}\Delta \varphi $ (blue dots) coupling disorder at the interface position $n=0$. In the first case, topological protection holds both interface states in the middle of the gap, while the dynamic disorder breaks chiral symmetry and spectrally splits the states. Robustness of the interface states against the static disorder, abruptly introduced at $m=50$, is illustrated for $\Delta \varphi =0.1\pi $ (b) and $\Delta \varphi =0.3\pi $ (d). In contrast, abruptly introduced dynamic perturbations with equivalent disorder amplitudes reveal a characteristic mode beating for the moderate disorder level (c) and a destruction of the localized modes for the stronger disorder level (e). The mode energy splitting of $\Delta E=0.248\pi $ at $\Delta \varphi =0.1\pi $(a) corresponds to a beating period of approximately 25 time steps (c). The coupling parameters used in all experiments are $\phi^{\prime}=0.1\pi $ and $\phi^{\prime\prime}=0.25\pi .$ }
\label{Fig5_PerturbedIntfcStates}
\end{figure}

\section{Topological protection against local disorder}
The number of interface states in each gap is conserved as long as chiral symmetry is preserved and no gap closure occurs. Moreover, if only a single state is situated in a gap its quasi-energy must stay in the middle required by the mirror-symmetry of the spectrum. Now, we check robustness of such symmetry protection by locally distorting one of the coupling parameters at the interface. 

First, we introduce a static disorder $\Delta \varphi $ to the coupler at the central interface position, namely $\phi _{0}^{m}=\phi^{\prime} +\Delta \varphi $. Numerically calculated spectra in the precense of such a disorder indicate that the interface states remains in the middle of the gap as expected [see black dotted line in Fig.\:\ref{Fig5_PerturbedIntfcStates}(a)], regardless of the disorder strength $\Delta \varphi $. Such perturbation is compatible with chiral symmetry and therefore the states are topologically protected against them. Indeed, in our experiments [see Figs.\:\ref{Fig5_PerturbedIntfcStates}(b) and (d)], where we introduce the coupling defect $\Delta \varphi $ abruptly as $\phi _{0}^{m}=\phi^{\prime} +\Delta \varphi $ for $m\geq50$, the trapped light remains largely intact, although some radiation is still shed away due to minor deformations of the localized modes’ profiles. This situation remains qualitatively unchanged even for the considerably large defect strength, applied in Fig.\:\ref{Fig5_PerturbedIntfcStates}(d).

In contrary, the field evolution completely changes if the original two round trip periodicity of the Floquet system is broken by an oscillating perturbation with a longer period as it is demonstrated in Figs.\:\ref{Fig5_PerturbedIntfcStates}(c) and (e). Still a local change of the coupling by $\Delta \varphi$ is induced at the interface for every second round trip, but with an alternating sign as $\phi _{0}^{m}=\phi^{\prime} +(-1)^m\Delta \varphi $ for $m\ge 50$. The resulting period doubling causes a back-folding of the otherwise unchanged energy spectrum of the bulk and projects both gaps including the energies of respective interface states onto one another. The resulting bulk back-folded spectrum is single band and therefore has trivial topology. Consequently, the interface states lose their topological protection, start to interact and leave their mid-gapped quasi-energy positions [see blue dotted lines in Fig.\:\ref{Fig5_PerturbedIntfcStates}(a)]. A rigorous analysis [see Appendix B for details] shows that chiral symmetry is indeed broken in this case. Note, that this is in contrast to the static SSH model, which can according to \cite{Fedorova19}  preserve a topological interface state exactly in the middle of the gap, leaving it topologically protected at least under slow (or fast) enough local periodic perturbations. For a moderate perturbation, which is again introduced after 100 round trips ($m\ge 50$), we experimentally detect in our system an additional slow beating [see Fig.\:\ref{Fig5_PerturbedIntfcStates}(c)] of the still existing, but now trivial eigenstates with slightly distinct eigenvalues. As topological protection is lost, the localized states can now freely dissolve in the continuum of the bulk, what indeed happens at stronger perturbation strengths, resulting in a complete dispersion of the guided light in the bulk domains [see Fig.\:\ref{Fig5_PerturbedIntfcStates}(e)]. The numerically obtained threshold for the destruction of interface states is around $\left| \Delta \varphi  \right|\approx 0.3\pi $ [see Fig.\:\ref{Fig5_PerturbedIntfcStates}(a)], while in the experiment the destruction took place for a slightly lower level of the perturbation $\left| \Delta \varphi  \right|\approx 0.25\pi $.

\section{Conclusions}
We have studied the dynamics of symmetry-protected chiral interface states in a one-dimensional Floquet photonic lattice in theory and experiment. A chiral symmetric discrete temporal system was realized by coupling two dissimilar fibre loops and monitoring the evolution of propagating pulses being subject to phase modulation and periodic variations of the coupling. We simultaneously observed two topologically protected localized states at a non-trivial interface between different bulk media and found them to be robust with respect to static perturbations obeying chiral symmetry. In contrast, non-stationary perturbations caused a period doubling break of topological protection and induced frequency shifts resulting in slow beating. For strong enough doubly periodic perturbations, we observed a complete destruction of the localized interface states.

\section{Acknowledgments}
This study was funded by the Deutsche Forschungsgemeinschaft in the framework of the IRTG 2101 “Guided light, tightly packed: novel concepts, components and applications” and the projects PE 523/18-1 and PE 523/14-1.

\section{Appendix A: Mode beating measurements}
Owing to a considerable spatial overlap of the two interface states with the single input pulse, launched at $n=0$ into loop $v$, both states are excited along with propagating bulk modes. After a few round trips, the latter ones have propagated away and the remaining light is trapped within the two localized stationary states. Since quasi-energies of these states are mismatched ($\Delta E=\pi $), they display a temporal beating within the overlap region. Concentrating on the power beat note in the short loop $u$, we observe the following superposition:
\begin{equation}
\begin{split}
   {{\left| u_{n,\exp }^{m} \right|}^{2}}=&W_{\exp }^{m}{{\left| Au_{n}^{0}+\sqrt{1-{{A}^{2}}}\,u_{n}^{\pi }{{e}^{im\Delta E}} \right|}^{2}} \\ 
  =&W_{\exp }^{m}\big[ {{A}^{2}}{{\left| u_{n}^{0} \right|}^{2}}+(1-{{A}^{2}}){{\left| u_{n}^{\pi } \right|}^{2}}\\
 +&2A\sqrt{1-{{A}^{2}}}\left| u_{n}^{0} \right|\left| u_{n}^{\pi } \right|\cos \left( m\Delta E \right) \big], 
\end{split}
\end{equation}
where ${{\left| u_{n,\exp }^{m} \right|}^{2}}$ is the experimentally measured energy profile at the interface and $W_{\exp }^{m}$ the total energy measured at step $m$ at the detector connected with loop $u$. ${{A}^{2}}$ and $1-{{A}^{2}}$ are energy portions of the two interface states. Their steady-state solutions with fixed normalized field profiles $u_{n}^{0}$ and $u_{n}^{\pi }$ are found numerically. Here we employ the fact that both field profiles are real valued due to the symmetry of the system and that the initial excitation was single site and could therefore not induce a relative phase between the states. By additionally evaluating the total energy $W_{\exp }^{m}$ which should stay constant in the ideal case we account for small varations of the net amplification in the loops. To estimate the energy portions $A^2$ and $1-{{A}^{2}}$ as displayed in Figs.\:4(d,e), we minimized the following function with respect to $A$:
\begin{equation}
\delta (A)=\sum\limits_{m=\Delta m}^{2\Delta m}\sum\limits_{n=-\Delta n}^{\Delta n}\beta_{n}^{m},\\
\end{equation}
where by definition
\begin{equation}
\begin{split}
&\beta_{n}^{m}\equiv\Big|{{\left[ {{\left| u_{n,\exp }^{m+1} \right|}^{2}}-W_{\exp }^{m+1}\left( {{A}^{2}}{{\left| u_{n}^{0} \right|}^{2}}+(1-{{A}^{2}}){{\left| u_{n}^{\pi } \right|}^{2}} \right) \right]}^{2}}\\
&-{{\left[ {{\left| u_{n,\exp }^{m} \right|}^{2}}-W_{\exp }^{m}\left( {{A}^{2}}{{\left| u_{n}^{0} \right|}^{2}}+(1-{{A}^{2}}){{\left| u_{n}^{\pi } \right|}^{2}} \right) \right]}^{2}} \Big|\\
&\div2A\sqrt{1-{{A}^{2}}}\left| u_{n}^{0} \right|\left| u_{n}^{\pi } \right|.
\end{split}
\end{equation}
In the ideal  case of topologically protected states the function $\delta (A)$ is equal or close to zero at the true value of $A$. The chosen time span of $\Delta m=30$ lays within the measurement window with high contrast beating and interference fringes at the interface. Importantly, the position span $\Delta n$ had been chosen within the typically narrow overlap region of the interface states, so that none of the summands in $\delta (A)$ diverge due to zero denominators. 

\section{Appendix B: Topological robustness of interface states against local stationary and non-stationary disorders}
Here, topological robustness of interface states against local stationary and non-stationary disorders of the coupling and phase parameters is analytically verified via consideration of fundamental symmetries, responsible for topological order of the lattice. To find out the influence of the coupling parameters ${{\phi }_{n}^\prime},{{\phi }_{n}^{\prime\prime}}$ and the phase shift parameters ${{\alpha }_{n}^\prime},{{\alpha }_{n}^{\prime\prime}}$ on topological order of the lattice, it is convenient to use the framework of Discrete Quantum Walks \cite{Kitagawa12,Kitagawa10a,Asboth13}. Namely, we represent an arbitrary state at a time step $m$ as a superposition of orthogonal locally defined two-component vectors (pseudospins)$\sum\limits_{n\in \mathbb{Z}}{{{\left( u_{n}^{m},v_{n}^{m} \right)}^{\text{T}}}\left| n \right\rangle }$.  The discrete evolution with the period of two round trips ($\Delta m=1$) is therefore governed by a unitary stroboscopic Floquet operator $\widehat{U}$, which is based on the model equations (1-4) of the main text: 
\begin{equation}
\widehat{U}({{\phi }_{n}^\prime},{{\phi }_{n}^{\prime\prime}} ,\overrightarrow{\alpha }^{\prime},\overrightarrow{\alpha }^{\prime\prime})\sum\limits_{n\in \mathbb{Z}}{\left( \begin{matrix}
   u_{n}^{m}  \\
   v_{n}^{m}  \\
\end{matrix} \right)}\left| n \right\rangle =\sum\limits_{n\in \mathbb{Z}}{\left( \begin{matrix}
   u_{n}^{m+1}  \\
   v_{n}^{m+1}  \\
\end{matrix} \right)\left| n \right\rangle },                                     
\end{equation}
where the Dirac “ket” vectors $\left| n \right\rangle $ denote eigenstates of the discrete position operator $\sum\limits_{n\in \mathbb{Z}}{\left| n \right\rangle }\left\langle  n \right|$,  ${{\phi }_{n}^\prime}$ and ${{\phi }_{n}^{\prime\prime}}$ are uniformly defined coupling parameters, applied in two sequential round trips, respectively, $\overrightarrow{\alpha }^{\prime}\equiv (..,{{\alpha }_{-1}^{\prime}},{{\alpha }_{0}^{\prime}},{{\alpha }_{1}^{\prime}},..,{{\alpha }_{n}^{\prime}},...)$ and $\overrightarrow{\alpha }^{\prime\prime}\equiv (..,{{\alpha }_{-1}^{\prime\prime}},{{\alpha }_{0}^{\prime\prime}},{{\alpha }_{1}^{\prime\prime}},..,{{\alpha }_{n}^{\prime,\prime}},...)$ are arrays of position-dependent phase shifts, applied in the long loop   in two sequential round trips (see the model equations (1-4) of the main text).  The topological interface has been previously introduced via $\left| {\alpha }^{\prime} \right|=\left| {\alpha }^{\prime\prime} \right|=\pi $ for $n\le 0$ and ${{\alpha }_{n}^\prime}={{\alpha }_{n}^{\prime\prime} }=0$, otherwise. Keeping the relation $\left| {\alpha }^{\prime} \right|=\left| {\alpha }^{\prime\prime} \right|$ we can parametrize the evolution operator in two ways as:   
\begin{equation}
   {{\widehat{U}}_{\pm }}=\sum\limits_{n\in \mathbb{Z}}{{{e}^{i{{\alpha }_{n}}\frac{1\pm 1}{2}}}{{\widehat{S}}_{e,n}}\widehat{B}(\pm {{\alpha }_{n}})\widehat{C}(\phi^{\prime\prime} )}\widehat{B}({{\alpha }_{n}}){{\widehat{S}}_{o,n}}\widehat{C}(\phi^{\prime}), 
\end{equation}
where by definition
\begin{align}
   \hat{B}(\alpha )&\equiv \left( \begin{matrix}
   {{e}^{-\frac{i\alpha }{2}}} & 0  \\
   0 & {{e}^{+\frac{i\alpha }{2}}}  \\
\end{matrix} \right),\\
   \hat{C}(\varphi )&\equiv \left( \begin{matrix}
   \cos \varphi  & i\sin \varphi   \\
   i\sin \varphi  & \cos \varphi   \\
\end{matrix} \right),  \\
   {{{\hat{S}}}_{e,n}}&\equiv \left( \begin{matrix}
   |n-1\rangle \langle n| & 0  \\
   0 & |n\rangle \langle n|  \\
\end{matrix} \right),\\
{{{\hat{S}}}_{o,n}}&\equiv \left( \begin{matrix}
   |n\rangle \langle n| & 0  \\
   0 & |n\rangle \langle n-1|  \\
\end{matrix} \right). 
\end{align}
Here the matrix $\hat{C}$ performs an on-site rotation of the pseudospin, depending on the respective coupling angle $\phi^{\prime}$ or $\phi^{\prime\prime}$. In contrast the matrices ${{\hat{S}}_{e,n}}$ and ${{\hat{S}}_{o,n}}$ just shift one of the two pseudospin components ($v$ and $u$, respectively) to a neighboring position, while leaving the remaining pseudospin component intact. The matrix $\hat{B}(\alpha )$ represents the phase shift applied to the $v$ loop, where we have extracted a global phase shift to symmetrize the expression.
Next, we perform a unitary rotation of the evolution operators corresponding to a time shift by half a round trip to represent the discrete walk in the so-called “chiral symmetric time frame” \cite{Asboth13}, where by definition the Floquet Hamiltonian obeys chiral symmetry
\begin{equation}
\begin{split}
&{{\widehat{U}}^{\prime }}_{\pm }=\widehat{C}(\phi^{\prime} /2){{\widehat{U}}_{\pm }}\widehat{C}(-\phi^{\prime} /2)\\
&=\sum\limits_{n\in \mathbb{Z}}{{{e}^{i\frac{{{\alpha }_{n}}(1\pm 1)}{2}}}\widehat{C}(\phi^{\prime} /2){{\widehat{S}}_{\text{e},n}}\widehat{B}(\pm {{\alpha }_{n}})\widehat{C}(\phi^{\prime\prime} )\widehat{B}({{\alpha }_{n}}){{\widehat{S}}_{o,n}}\widehat{C}(\phi^{\prime} /2)}.                                                                                                                       
\end{split}
\end{equation}
Here the prime indicates the new time frame. Next, we introduce on-site perturbations of the coupling parameter $\phi^{\prime}$  for odd round trips, in two different ways, namely:
\begin{equation}
\begin{split}
  &{{\widehat{U}}^{\prime }}_{\pm ,j}=\sum\limits_{n\in \mathbb{Z}}{{e}^{i\frac{{{\alpha }_{n}}(1\pm 1)}{2}}}\widehat{C}((\phi^{\prime} +{{(-1)}^{j-1}}\Delta {{\varphi }_{n}})/2)\cdot\\
 &\cdot {\widehat{S}}_{\text{e},n}\widehat{B}(\pm {{\alpha }_{n}})\widehat{C}(\phi^{\prime\prime} )\widehat{B}({{\alpha }_{n}}){{\widehat{S}}_{o,n}}\widehat{C}((\phi^{\prime} +\Delta {{\varphi }_{n}})/2),\\
 &\quad j=1  \rm{\: or \:}2, \Delta {{\varphi }_{n}}=\left\{ \begin{matrix}
   0,  \quad  n\ne 0,  \\
   \Delta \varphi , n=0.  \\
\end{matrix} \right.   
\end{split}
\end{equation}
The first perturbed evolution operator ($j=1$) corresponds to the experimentally investigated case of a static perturbation, introduced at the interface position $n=0$ in the main text. The second evolution operator ($j=2$) describes one half of the dynamic perturbation with periodically alternating sign ($\Delta m=2$), that has been introduced in the main text as well. In order to embrace the full period of this perturbation, one would have to include one more time step into consideration, however the current instantaneous picture within the time window $\Delta m=1$ is already enough to argue chiral symmetry breaking.
Indeed, to verify topological protection, we define the fundamental chiral ${{\widehat{T}}_{1}}$ and time-reversal ${{\widehat{T}}_{2}}$ symmetries in the above introduced time frame as follows:
\begin{align}
   &{{\widehat{T}}_{1}}={{\widehat{\sigma }}_{y}}\otimes \sum\limits_{n}{\left| n \right\rangle }\left\langle  n \right|=\oint\limits_{\text{BZ}}{\left| k \right\rangle \left\langle  k \right|dk}\otimes {{\widehat{\sigma }}_{y}}, \\ 
  &{{\widehat{T}}_{2}}={{\widehat{\sigma }}_{x}}K\otimes \sum\limits_{n}{\left| n \right\rangle }\left\langle  n \right|=\oint\limits_{\text{BZ}}{\left| -k \right\rangle \left\langle  k \right|dk}\otimes {{\widehat{\sigma }}_{x}}K, 
\end{align}
where ${{\widehat{\sigma }}_{x,y}}$ are standard unitary Pauli matrices, $K$ is the complex conjugate operator. Note that $\widehat{T}_{2}^{2}=+1,$ thus revealing an integer spin statistics \cite{Schnyder08} of the discrete pseudospin walk. The effective Floquet Hamiltonian ${{\widehat{H}}^{\prime }}$ is said to obey time-reversal (chiral) symmetry, if it commutes (anticommutes) with the associated operator:
\begin{align}
{{\widehat{T}}_{1,2}}{{\widehat{H}}^{\prime }}\widehat{T}_{1,2}^{-1}=\mp {{\widehat{H}}^{\prime }}                                                        
\end{align}
or, by doing Taylor expansion of the matrix exponential ${{\widehat{U}}^{\prime }}=\exp (i{{\widehat{H}}^{\prime }})$, we equivalently obtain the relations
\begin{align}  
{{\widehat{T}}_{1,2}}{{\widehat{U}}^{\prime }}{\widehat{T}_{1,2}}^{-1}={{\widehat{U'}}}^{-1}    \Leftrightarrow     {{\widehat{T}}_{1,2}}{{\widehat{H}}^{\prime }}{\widehat{T}_{1,2}}^{-1}=\mp {{\widehat{H}}^{\prime }},        
\end{align}     
which can be directly applied to the Floquet evolution operators by taking
\begin{equation}
\begin{split}
  \widehat{T}{{\widehat{U'}}}_{\pm ,j}\widehat{T}_{{}}^{-1}=&\sum\limits_{n\in \mathbb{Z}}\Big({{\widehat{T}}_{{}}}\widehat{C}\left( (\phi^{\prime} +{{(-1)}^{j-1}}\Delta {{\varphi }_{n}})/2 \right)\widehat{T}_{{}}^{-1}\\
  &\cdot{{\widehat{T}}_{{}}}{{\widehat{S}}_{\text{e},n}}\widehat{T}_{{}}^{-1}{{\widehat{T}}_{{}}}{{e}^{i\frac{{{\alpha }_{n}}(1\pm 1)}{2}}}\widehat{B}(\pm {{\alpha }_{n}})\widehat{C}(\phi^{\prime\prime} )\widehat{B}({{\alpha }_{n}})\widehat{T}_{{}}^{-1}  \\ 
 & \cdot {{\widehat{T}}_{{}}}{{\widehat{S}}_{o,n}}\widehat{T}_{1,2}^{-1}{{\widehat{T}}_{{}}}\widehat{C}\left( (\phi^{\prime} +\Delta {{\varphi }_{n}})/2 \right)\widehat{T}_{{}}^{-1}\Big), 
  \end{split}
\end{equation}
 and
\begin{equation}
\begin{split}
  &{{\widehat{U'}}}_{\pm ,j}^{-1}=\sum\limits_{n\in \mathbb{Z}}{{\widehat{C}}^{-1}}\left( (\phi^{\prime} +\Delta {{\varphi }_{n}})/2 \right)\widehat{S}_{\text{o},n}^{-1}\cdot\\
  &\cdot{{\left( {{e}^{i\frac{{{\alpha }_{n}}(1\pm 1)}{2}}}\widehat{B}(\pm {{\alpha }_{n}})\widehat{C}(\phi^{\prime\prime} )\widehat{B}({{\alpha }_{n}}) \right)}^{-1}}\cdot\\
  &\cdot\widehat{S}_{\text{e},n}^{-1}{{\widehat{C}}^{-1}}\left(\dfrac{\phi^{\prime} +{{(-1)}^{j-1}}\Delta {{\varphi }_{n}}}{2} \right), 
\end{split}
\end{equation}
where the indices of ${{\widehat{T}}_{1,2}}$ are omitted for a better readability. Since the system parameters $\phi^{\prime} ,\phi^{\prime\prime} ,\Delta \varphi $ and ${{\alpha }_{n}}$ are mutually independent, the symmetry conditions for ${{\widehat{U}}^{\prime }}_{\pm }$ are fulfilled, if and only if these conditions are respected by its constituents, namely if 
\begin{align}
   {{\widehat{T}}_{1,2}}{{{\hat{S}}}_{\text{o},n}}\widehat{T}_{1,2}^{-1}=\hat{S}_{\text{e}\text{,}n}^{-1},  \forall n,
\end{align}
as well as
\begin{align}
  {{\widehat{T}}_{1,2}}\widehat{C}\left( (\phi^{\prime} +{{(-1)}^{j-1}}\Delta \varphi )/2 \right)\widehat{T}_{1,2}^{-1}={{\widehat{C}}^{-1}}\left( (\phi^{\prime} +\Delta \varphi )/2 \right),  
\end{align}
at $n=0,j=1, 2$ and if 
\begin{align}  
\label{eq29} 
 &{{\widehat{T}}_{i}}{{e}^{i{{\alpha }_{n}}}}\widehat{B}({{\alpha }_{n}})\widehat{C}\left( \phi^{\prime\prime}  \right)\widehat{B}({{\alpha }_{n}})\widehat{T}_{1,2}^{-1}={{({{e}^{i{{\alpha }_{n}}}}\widehat{B}({{\alpha }_{n}})\widehat{C}\left( \phi^{\prime\prime}  \right)\widehat{B}({{\alpha }_{n}}))}^{-1}},\\
 \label{eq30} 
 &{{\widehat{T}}_{i}}\widehat{B}(-{{\alpha }_{n}})\widehat{C}\left( \phi^{\prime\prime}  \right)\widehat{B}({{\alpha }_{n}})\widehat{T}_{1,2}^{-1}={{(\widehat{B}(-{{\alpha }_{n}})\widehat{C}\left( \phi^{\prime\prime}  \right)\widehat{B}({{\alpha }_{n}}))}^{-1}}
\end{align}             
for all $n$ and for $i=1,2$. The upper \eqref{eq29} and the lower \eqref{eq30} equation is derived for ${{\widehat{U}}^{\prime }}_{+}$ and ${{\widehat{U}}^{\prime }}_{-}$, respectively.

One can rigorously check that the first condition is satisfied for both time-reversal and chiral symmetry operators. The second condition is satisfied for both symmetries, only if $\phi^{\prime} +{{(-1)}^{j-1}}\Delta \varphi =\phi^{\prime} +\Delta \varphi +2\pi l$, where $l$ is an arbitrary integer number. It implies that the static local disorder respects both chiral and time-reversal symmetry, while the dynamic local disorder generally breaks them. To finally prove that the latter perturbation breaks topological order of the system, one has to additionally show that so-called particle-hole symmetry (PHS), defined as ${{\widehat{T}}_{3}}={{\widehat{T}}_{2}}{{\widehat{T}}_{1}}=-{{\widehat{\sigma }}_{z}}K\otimes \sum{\left| n \right\rangle \left\langle  n \right|}$, is broken as well. Rigorous calculations show that this is indeed the case and thus, in accordance with the periodic classification of topological insulators \cite{Schnyder08}, the topological order does not protect the interface states against such dynamic perturbations, in agreement with the numerical and experimental results, presented in the main text in Figs.\:5(a) and 5(c,e), respectively. Finally, the last two conditions (\ref{eq29}) and (\ref{eq30}) can be transformed into the expressions
\begin{align}
 &\left\{ \begin{matrix}
   {{\widehat{T}}_{1}}{{e}^{i{{\alpha }_{n}}}}\ldots\widehat{T}_{1}^{-1}-{{({{e}^{i{{\alpha }_{n}}}}\ldots)}^{-1}}&=&\widehat{f}(\phi^{\prime\prime} ,{{\alpha }_{n}})\cdot \sin {{\alpha }_{n}}  \\
   {{\widehat{T}}_{2}}{{e}^{i{{\alpha }_{n}}}}\ldots\widehat{T}_{2}^{-1}-{{({{e}^{i{{\alpha }_{n}}}}\ldots)}^{-1}}&=&\widehat{g}(\phi^{\prime\prime} ,{{\alpha }_{n}})\cdot \sin {{\alpha }_{n}}\cos \phi^{\prime\prime}   \\
\end{matrix}, \right. \\ 
&\left\{ \begin{matrix}
   {{\widehat{T}}_{1}}\ldots\widehat{T}_{1}^{-1}-{(\ldots)}^{-1}&=&{{\widehat{\sigma }}_{y}}2i\sin {{\alpha }_{n}}\sin \phi^{\prime\prime}   \\
   {{\widehat{T}}_{2}}\ldots\widehat{T}_{2}^{-1}-{{(\ldots)}^{-1}}&=&0  \\
\end{matrix},\right. 
\end{align}
$\forall n$ for ${{\widehat{U}}^{\prime }}_{+}$ and ${{\widehat{U}}^{\prime }}_{-}$, respectively. $\widehat{f}(\phi^{\prime\prime} ,{{\alpha }_{n}})$ and $\widehat{g}(\phi^{\prime\prime} ,{{\alpha }_{n}})$ are non-zero matrix functions, whose exact forms are not important here. One can infer that any value of ${{\alpha }_{n}}$ except 0 and $\pi$ breaks chiral (chiral and time-reversal) symmetry for the parametric continuation ${{\widehat{U}}^{\prime }}_{-}$ (${{\widehat{U}}^{\prime }}_{+}$). In case of ${{\widehat{U}}^{\prime }}_{-}$, any arbitrary phase shift except 0 and $\pi$ breaks chiral symmetry and therefore the system becomes topologically trivial \cite{Schnyder08}. In case of ${{\widehat{U}}^{\prime }}_{+}$, all symmetries brake simultaneously and the system becomes trivial as well. It is worth mentioning however, that in the latter case chiral symmetry gets broken exclusively due to the global phase shifts ${{e}^{i{{\alpha }_{n}}}}$, implying that an experimental arrangement with the anti-symmetric phase modulation in both $u$ and $v$ loops should display topologically protected states even for arbitrary values of ${{\alpha }_{n}}$. In conclusion, the exclusive choice of 0 and $\pi$ for the phase modulation parameter in our experiments is motivated by the preserved topological order of the walk. 
                     
\printbibliography

\end{document}